\documentclass[%
 reprint,
 amsmath,amssymb,
 aps,
]{revtex4-1}

\usepackage{graphicx}
\usepackage{dcolumn}
\usepackage{bm}

\begin{document}


\title{Evidence for a Solenoid Phase of Supercoiled DNA}

\author{Andrew Dittmore}
\author{Keir C. Neuman}
\affiliation{Laboratory of Single Molecule Biophysics, National Heart, Lung, and Blood Institute, National Institutes of Health, Bethesda, Maryland 20892, USA}

\date{\today}

\begin{abstract}
In mechanical manipulation experiments, a single DNA molecule overwound at constant force undergoes a discontinuous drop in extension as it buckles and forms a superhelical loop (a plectoneme). Further overwinding the DNA, we observe an unanticipated cascade of highly regular discontinuous extension changes associated with stepwise plectoneme lengthening. This phenomenon is consistent with a model in which the force-extended DNA forms barriers to plectoneme lengthening caused by topological writhe. Furthermore, accounting for writhe in a fluctuating solenoid gives an improved description of the measured force-dependent effective torsional modulus of DNA, providing a reliable formula to estimate DNA torque. Our data and model thus provide context for further measurements and theories that capture the structures and mechanics of supercoiled biopolymers.
\end{abstract}
\maketitle

\begin{figure}\label{fig1}
\includegraphics{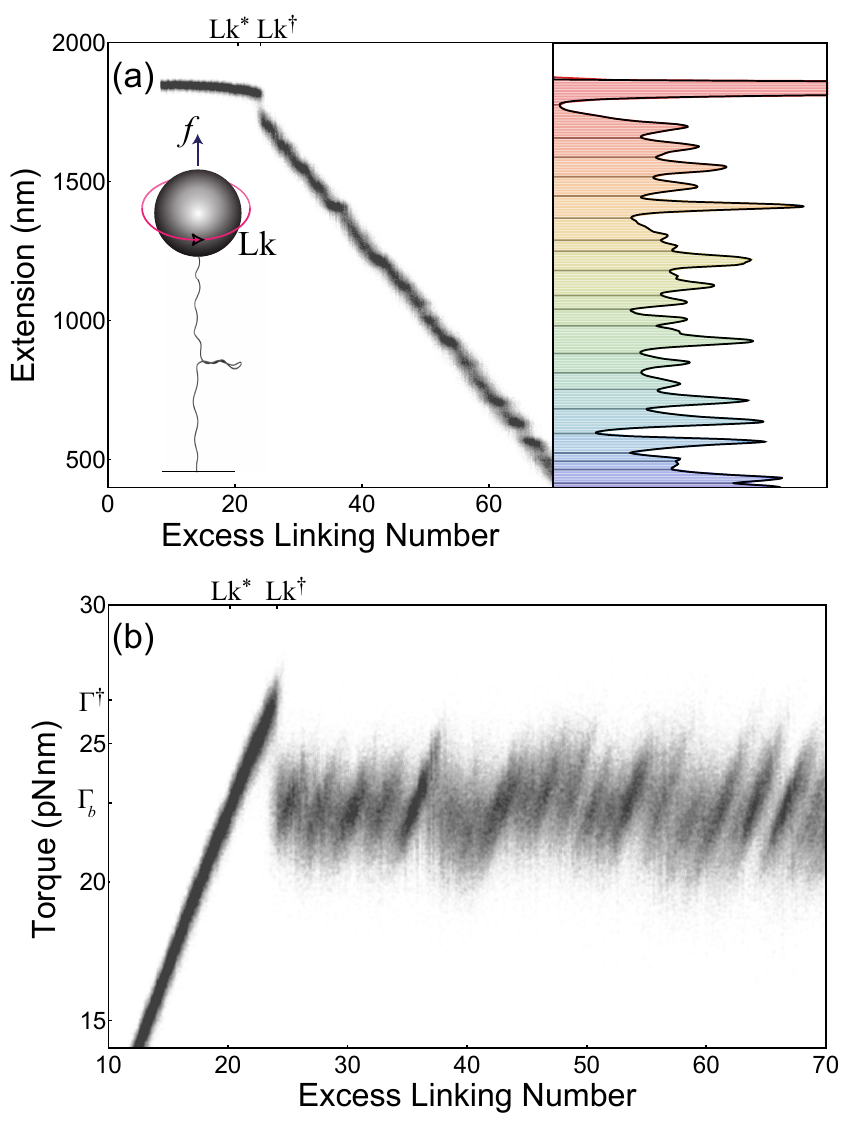}
\caption{Single-molecule data on $2~\mu\text{m}$ DNA at 3.6~pN force and 0.5~M salt. (a) End-to-end extension of the DNA vs excess linking number \cite{strick}. The schematic depicts the surface-immobilized DNA molecule overwound by ${\rm Lk}$ rotations of the tethered magnetic bead at constant upward force $f$. The horizontal buckled structure is a {\it plectoneme}. Adjacent points recorded at 5-ms intervals are separated by $10^{-4}$ in ${\rm Lk}$. The extension histogram (right; 10-nm bins) reveals a series of discrete post-buckling states colored in varying hues. The scale is from 0 to 8000. (b) Torque vs excess linking number \cite{note1}. The torsional response is linearly elastic up to the maximum torque $\Gamma^\dagger$ at the point of buckling (${\rm Lk}^\dagger$, near ${\rm Lk}=24$). The torque then varies about an average value $\Gamma_b$ (${\rm Lk>{\rm Lk}^\dagger}$), increasing linearly in ${\rm Lk}$ within each discrete state while fluctuating discontinuously among states. The {\it torque overshoot}, ($\Gamma^\dagger-\Gamma_b$) assumed to occur from ${\rm Lk}^*$ to ${\rm Lk}^\dagger$, is comparable in scale.}
\end{figure}

\begin{figure}\label{fig2}
\includegraphics{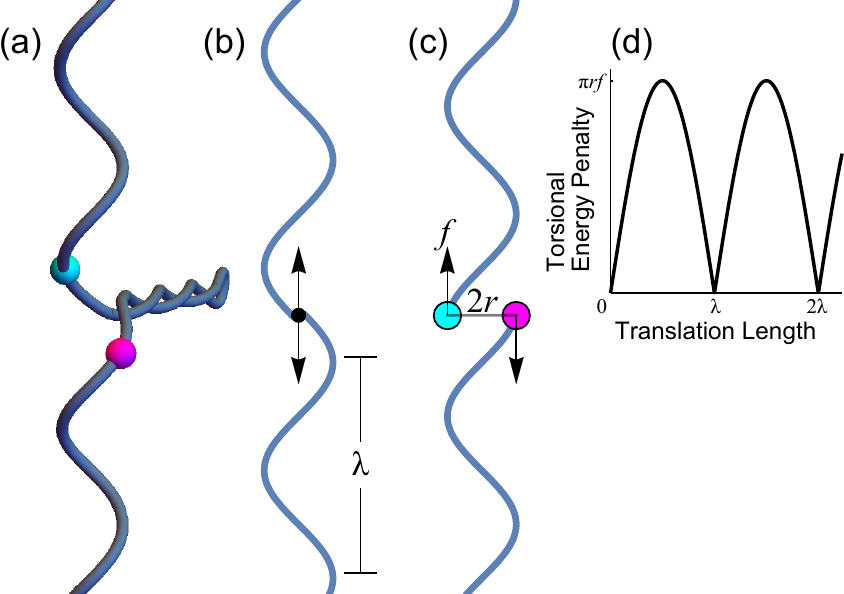}
\caption{Schematic structural model of plectoneme lengthening in discrete steps. (a) Plectoneme joining a solenoid. (b) Planar representation of solenoid of wavelength $\lambda$. The plectoneme projects into the plane at the black dot, where up and down arrows are internal forces in balance at this point. (c) Translational symmetry breaking. Translating either end of the solenoid by $\lambda/4$ to lengthen the plectoneme (total translation length $\lambda/2$) would incur a maximum torsional energy penalty due to the force-couple, $2rf$ (with $r\sim(\beta f)^{-1}$), integrated through rotation by $\pi/2$. This estimate is used only to establish the \emph{existence} of periodic energy barriers exceeding $\beta^{-1}$ and ignores deformation or other energy contributions to the unknown transition state structure. Colored points correspond to initial contour positions in a. (d) Torsional energy penalty vs translation length. The symmetry breaking depicted in c results in periodic barriers to continuous plectoneme lengthening. The symmetry in b is reflected in positions with zero energy penalty that repeat periodically in incremental translation lengths of one solenoid wavelength, predicting a step size \cite{SI}.}
\end{figure}

\begin{figure}\label{fig3}
\includegraphics{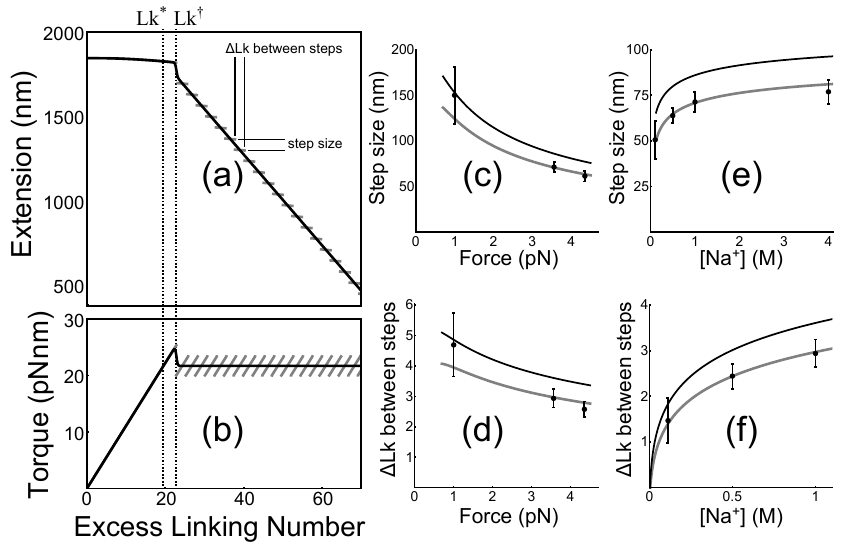}
\caption{Model and comparison with data as a function of force and salt concentration. (a) End-to-end extension vs excess linking number calculated under conditions identical to Fig.~1a. The average extension curve (eqn.~\ref{eq:eqn5}) is plotted in black. The step size and difference in linking number between steps calculated from the solenoid model are indicated on the discontinuous gray curve. (b) Torque vs excess linking number calculated under conditions identical to Fig.~1b. The average torque curve (eqn.~\ref{eq:eqn6}) is plotted in black. The torque increases linearly in each state in the discontinuous gray curve. (c) Step size vs force and (d) change in linking number between steps vs force. (e) Step size vs salt concentration and (f) change in linking number between steps vs salt concentration. The black curves correspond to a unit wavelength of solenoid, equal to $\langle{X_u(f,L_0,{\rm Lk}^\dagger)}\rangle/{\rm Lk}^\dagger$. The gray curves correct for $\sqrt{A/(\beta f)}$ bending fluctuations \cite{moroz}, which are expected to shorten each adjacent half-wave of solenoid above and below the plectoneme. Error bars are standard error of the mean across multiple post-buckling steps (Fig.~1). Force was varied at fixed 1~M monovalent salt; and salt concentration was varied at fixed 3.6~pN force.}
\end{figure}

\begin{figure}\label{fig4}
\includegraphics[width=2.7 in]{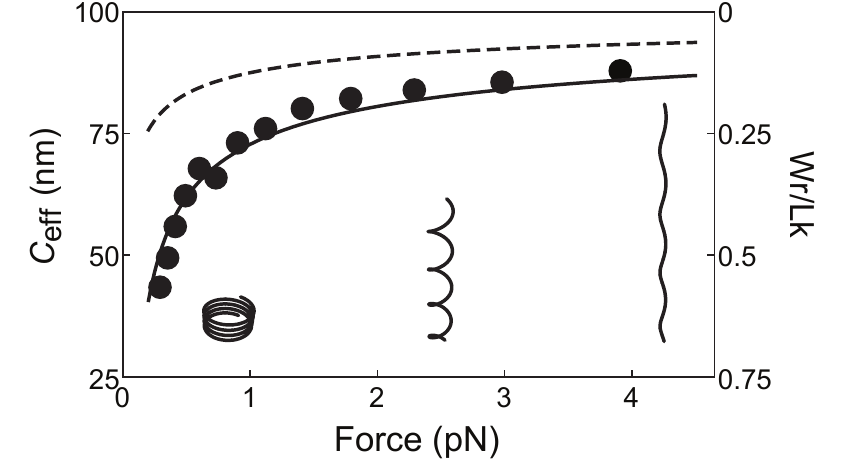}
\caption{Force dependence of DNA torsional elasticity and partitioning of twist and writhe (see eqn.~\ref{eq:eqn1}). The effective torsional persistence length $C_{\rm eff}$ increases with force and asymptotically approaches $C$. Points are data from Mosconi {\it et al.} \cite{mosconi}. The dashed curve is the theoretical prediction of Moroz and Nelson \cite{moroz}, $C_{\rm eff}(f)\approx(\frac{1}{C}+\frac{1}{4A\sqrt{\beta Af}})^{-1}$. The solid curve (eqn.~\ref{eq:eqn7}), $C_{\rm eff}(f)\approx(\frac{1}{C}+\frac{1}{4A\sqrt{\beta Af}})^{-1}-C(1-x_{\rm WLC})$, gives an improved description of data by including an estimate of writhe per link in a solenoid, ${\rm Wr}/{\rm Lk}\approx1-x_{\rm WLC}$ \cite{note5}. Both curves are plotted with bending persistence length $A=50~\text{nm}$ and bare twist persistence length $C=100~\text{nm}$. This comparison indicates that the total writhe per link (${\rm Wr}/{\rm Lk}=1-{\rm Tw}/{\rm Lk}$; right axis) is consistent with the sum of a solenoid (depicted schematically) plus fluctuations about this shape \cite{note6}.}
\end{figure}

Supercoiling of DNA permits its compaction in a cell and is crucial to the biological activities of DNA interacting proteins. Understanding the structures and mechanics of DNA supercoiling is therefore a topic of fundamental interest \cite{benham}. Experimentally, the force and linking number of a topologically closed DNA double helix can be specified in a single-molecule manipulation experiment with torsional control \cite{strick}. Following a theorem that states that the imposed torsion or excess linking number ({\rm Lk}) partitions into interconverting components of twist ({\rm Tw}) and writhe ({\rm Wr}) \cite{calugareanu,white,fuller}, the DNA molecule can either twist about its backbone or writhe into more complex shapes that depend on force and solution conditions. These DNA shapes produce mechanical signatures observable by experiment.

At a critical torque, twisted DNA buckles: It loops upon itself and wraps into a superhelical writhed structure called a plectoneme (Fig.~1). The plectoneme lengthens with additional imposed turns. Although this is widely assumed to be a smooth and continuous process, here we present experimental evidence that a DNA plectoneme lengthens in discrete steps that occur with a periodicity that exceeds the characteristic plectoneme repeat length (Fig.~1a). This behavior indicates supercoiling involves unexpected periodic free-energy barriers of unknown origin.

The observed plectoneme lengthening steps can be interpreted as discontinuous exchanges of twist for writhe. At fixed force and torsion, the conjugate variables, extension and torque, are free to fluctuate. Since torque increases with increasing twist, the discrete plectoneme lengthening phenomenon corresponds to a progressive buildup, then release, of DNA torque (Fig.~1b).

To explain this phenomenon, we propose a model in which barriers to plectoneme lengthening arise from the force-extended DNA adopting a solenoid structure (Fig.~2). If the force-extended region were straight without writhe, it would have continuous translational symmetry and adding length to the plectoneme would increase energy smoothly. In contrast, a solenoid shape is periodic and has discrete translational symmetry, resulting in a periodic energy associated with adding length to the plectoneme: To preserve boundary matching conditions in the joining region, the solenoid discontinuously feeds into the plectoneme in steps of one solenoid wavelength (Fig.~2). 
To compare this model with measurements, we first develop a heuristic computational framework describing equilibrium DNA mechanics (eqns.~\ref{eq:eqn1}-\ref{eq:eqn6}) \cite{SI,note4}, and then use this to calculate steps predicted by a solenoid model (Fig.~3); we find that measurements are broadly consistent with predictions. In further support of this model, we find that changes in torsional elasticity with force can be accounted for by estimating the writhe of the solenoid (Fig.~4). Thus, the solenoid model we propose to explain our primary observation of an unanticipated discontinuous supercoiling phenomenon potentially also resolves an existing discrepancy between theory and experiment regarding the decrease in the effective torsional persistence length with decreasing force (eqn.~\ref{eq:eqn7}).

Solenoid shapes were anticipated by theorists \cite{plectenoid}, and expected to range from a circularly writhed coil at zero force to a straight-twisted rod at high force (Fig.~4 inset). In their treatment of DNA supercoiling, this intuition was expressed by Marko and Siggia \cite{plectenoid}: ``... whatever extension exists must be bridged by a helical random coil or solenoid.'' Though this structure has not been seen in DNA, solenoids directly visualized in actin filaments may have biological functions \cite{leijnse}. However, a straight-shape hypothesis has prevailed in the DNA literature, in part because thermally fluctuating shapes are difficult to visualize, and also because calculations assuming an initially straight elastic rod show that the lowest energy structure is a straight-twisted rod and not a solenoid \cite{fain}. We can conclude from this that if solenoids occur in DNA, as they do on a larger scale in microscopic actin filaments, these solenoids are likely different than the solenoids of classical elastic rod theory \cite{love}.

Reconciliation with theory comes by relaxing the assumption that the DNA is initially straight. In the presence of intrinsic sequence-dependent bends as well as entropic bending fluctuations \cite{okonogi}, overwound DNA is never straight, and can reach a lower energy by trading twist for writhe \cite{moroz}. Consistent with this, Thompson and Champneys showed that torsion causes initially bent elastic rods to develop a nonclassical, one-turn-per-wave solenoid \cite{thompson}. This previously unrecognized solenoid is not a new solution to the elastic rod equations, but is instead the more general form of the straight-twisted rod solution \cite{champneys}, in agreement with intuition \cite{plectenoid}.

Motivated to investigate whether signatures of a one-turn-per-wave solenoid can be found in the mechanical characteristics of overwound DNA, we develop a heuristic computational approach to reproduce the observed behavior of supercoiled DNA \cite{SI,note4}. To compare our model with data, we first calculate the continuous or average elastic response curves, and then include the predicted discrete states that correspond to twist-writhe fluctuations and plectoneme lengthening steps that arise due to the solenoid structure adopted by the extended DNA.

Using three phenomenological constants, ${\rm c_1}$, ${\rm c_2}$, and ${\rm c_3}$, we develop a heuristic quantitative model for the elastic response of DNA under tension and torsion, favoring empirical relationships where current theory is lacking. We limit our attention to long molecules of positively supercoiled DNA, subject to small forces that allow thermal bending fluctuations, and small torques that preserve the rigidly stacked double-helical structure \cite{SI,note4}. Under these conditions, DNA is the prototypical example of a semiflexible and twist-storing polymer, well described as an elastic rod with bend persistence length $A=50~\text{nm}$ and twist persistence length $C=100~\text{nm}$ much larger than the constituent chemical monomers. These length scales are related to the distances over which bend and twist fluctuations induced by the thermal energy, $\beta^{-1}=4.1~\text{pN}\,\text{nm}$, become uncorrelated. For a DNA molecule of contour length $L_0$, imposed torsion increases the interlinking of the two strands of the double helix by excess linking number ${\rm Lk}$. The force $f$ and concentration $c$ of monovalent salt are held constant as the DNA is overwound.

The torque $\Gamma$ increases in direct proportion to torsion ${\rm Lk}$ according to the linear elasticity equation,
\begin{subequations}\label{eq:eqn1}
\begin{align}
\langle\Gamma_u(f,L_0,{\rm Lk})\rangle&=2\pi C_{\rm eff}(f)/(\beta L_0){\rm Lk}\\
&=2\pi C/(\beta L_0)\langle{{\rm Tw}}\rangle,
\end{align}
\end{subequations}
where braces denote the thermal average. 
With ${\rm Lk}={\rm Tw}+{\rm Wr}$, it is clear that the conventional use of the effective torsional modulus (force-dependent $C_{\rm eff}/\beta$ in eqn.~\ref{eq:eqn1}a) instead of the microscopic twist modulus (force-independent $C/\beta$ in eqn.~\ref{eq:eqn1}b) is a concise statement that overwound DNA develops writhe (see Fig.~4).

The nonzero writhe is evidence for bends adopting chirality in the direction of torque \cite{moroz}, leading to a reduction of the extension of unbuckled DNA. We start with the Marko-Siggia wormlike chain (WLC) model to obtain the normalized extension $x_{\rm{WLC}}(f)$ of a length $L_0$ DNA molecule at zero torque \cite{marko}, and note that an increase in writhe due to torque-enhanced fluctuations explains the subtle, approximately quadratic, decrease in the extension vs linking number curve prior to buckling (Fig.~1). We observe that a fraction $1/{\rm c_1}$ of the torsional energy reduction relative to a straight-twisted rod goes into the work of decreasing the extension at constant force. With $\rm{c_1}\approx3$ a constant,
\begin{equation}\label{eq:eqn2}
\langle{X_u(f,L_0,{\rm Lk})}\rangle\approx{L_0x_{\rm WLC}(f)-\frac{2(\pi{\rm Lk})^2(C-C_{\rm eff}(f))}{{\rm c_1}\beta L_0 f}}
\end{equation}
is a simple approximation for the extension vs excess linking number behavior of twisted and writhed DNA prior to buckling. Once the DNA buckles, its end-to-end extension decreases linearly, on average, due increases in linking number being taken up as writhe in the plectoneme. Each additional turn lengthens the plectoneme and shortens the force-extended portion by $q\approx3\pi \Gamma_b x_{\rm WLC}/(2 f)$ \cite{clauvelin}, where
\begin{equation}\label{eq:eqn3}
\Gamma_b(f,c)\approx{\rm c_2}f^{0.72}/\ln(c)
\end{equation}
is the average torque in buckled DNA. With monovalent salt concentration $c$ in $\text{mM}$ and force $f$ in $\text{pN}$, the constant $\rm{c_2}\approx{54}$. The $\ln(c)^{-1}$ scaling is expected from electrostatic models \cite{debye}, whereas the $f^{0.72}$ power-law is empirical \cite{mosconi}, and not yet explained by theory.

Next, comparing eqns.~\ref{eq:eqn1} and ~\ref{eq:eqn3}, we locate a useful feature point of DNA supercoiling curves: The torque in unbuckled DNA is equal to $\Gamma_b$ at ${\rm Lk}^*=\beta L_0\Gamma_b/(2\pi C_{\rm eff})$. This is assumed to be the point at which the post-buckling slope intersects the $\langle{X_u}\rangle$ vs ${\rm Lk}$ curve (eqn.~\ref{eq:eqn2}). By extrapolating the slope $q$ from ${\rm Lk^*}$, we can write the post-buckling extension as
\begin{equation}\label{eq:eqn4}
\langle{X_b(f,c,L_0,{\rm Lk})}\rangle=\langle{X_u(f,L_0,{\rm Lk}^*)}\rangle-q({\rm Lk}-{\rm Lk}^*).
\end{equation}

Finally, to complete the average elastic response curves, we estimate details near the buckling transition. The DNA extension at buckling is marked by a discontinuous transition between unbuckled (eqn.~\ref{eq:eqn2}) and buckled (eqn.~\ref{eq:eqn4}) states. These are occupied with equal probability at the transition midpoint ${\rm Lk}^\dagger$, where unbuckled DNA and buckled DNA featuring a small plectoneme are equal in energy. We assume the torque increase from ${\rm Lk}^*$ to ${\rm Lk}^\dagger$ is proportional to this linking number difference (eqn.~\ref{eq:eqn1}), with $({\rm Lk}^\dagger-{\rm Lk}^*)\approx{\rm c_3}f/\Gamma_b(f,c)$ and constant $\rm{c_3}\approx20~\text{nm}$ used to approximate the scale of this feature for DNA \cite{note2}, which is known as the {\it torque overshoot} (Fig.~1). Near the discontinuity, the microscopic nature of this transition (energy scale comparable to $\beta^{-1}$) permits coexistence fluctuations that give rise to smoothly varying average extension and torque curves. Buckling is dictated by the exchange of free energy and described using Boltzmann statistics, with probabilities $P_u=(1+\exp{[4\pi^2({\rm Lk}^\dagger-{\rm Lk}^*)({\rm Lk}-{\rm Lk}^\dagger)C_{\rm eff}/L_0])}^{-1}$ of the unbuckled state and $P_b=1-P_u$ of the buckled state \cite{note3,SI}. Joining eqns.~\ref{eq:eqn2} and \ref{eq:eqn4},
\begin{equation}\label{eq:eqn5}
\langle{X(f,c,L_0,{\rm Lk})}\rangle=P_u\langle{X_u}\rangle+P_b\langle{X_b}\rangle.
\end{equation}
Similarly, joining eqns.~\ref{eq:eqn1} and \ref{eq:eqn3},
\begin{equation}\label{eq:eqn6}
\langle{\Gamma(f,c,L_0,{\rm Lk})}\rangle=P_u\langle{\Gamma_u}\rangle+P_b\langle{\Gamma_b}\rangle.
\end{equation}

Eqns.~\ref{eq:eqn5} and \ref{eq:eqn6} describe the \emph{average} elastic response to supercoiling a DNA molecule whose total length is partitioned into a plectonemic fraction and a force-extended fraction. In general, lengths of DNA can exchange between these fractions, interconverting twist and writhe, leading to variance and discontinuities in $X$ and $\Gamma$ at a given value of ${\rm Lk}$ (Fig.~1). The pronounced discontinuity at the buckling transition is attributed to plectoneme formation. We find that as {\rm Lk} is increased past the critical point ${\rm Lk}^\dagger$ at buckling, a series of discontinuities ensues (Fig.~1).

We investigate whether a structural model might explain this. In the most general sense, the phenomenon we observe might be explained by the energy of locally bent DNA (the {\it tails}) feeding into the plectoneme. Linking-number-dependent variations in the tails could occur due to superstructure or sequence causing variations in bending energy. Since we observe modulations on what is otherwise a regular, periodic signal (Fig.~1), we attribute the periodicity to superstructure and the modulations to sequence. Since the simplest periodic superstructure is a solenoid, we consider whether the existence of a force-extended solenoid is consistent with the observed elasticity and supercoiling behaviors of DNA. We do this through calculations of a solenoid model, which we compare to data without fitting.

Marko and Siggia anticipated a ``plecto-noid'' coexistence state of torsionally buckled DNA \cite{plectenoid}, consistent with our proposed structural model (Fig.~2). Our phenomenological framework for constructing the average elastic response curves (continuous black curves in Fig.~3a) provides all necessary parameters for predicting discrete states in the context of this model \cite{SI}, where writhe in extended DNA causes discontinuous plectoneme lengthening corresponding to an extension change, or step size, of one solenoid wavelength (Fig.~2). In the simplest case of a one-turn-per-wave solenoid, the expected number of waves is equal to the imposed number of turns, ${\rm Lk}$. Therefore, absent fluctuations in a simple solenoid shape (Fig.~2), the expected step size is equal to the extension per wave, or wavelength, $\langle{X_u(f,L_0,{\rm Lk}^\dagger)}\rangle/{\rm Lk}^\dagger$, at buckling (Fig.~3). A more refined estimate accounts for bending fluctuations \cite{moroz}, which are expected to facilitate barrier crossing and reduce the predicted step size by $2\sqrt{A/(\beta f)}$ (Fig.~3). The corresponding linking number change $\Delta{\rm Lk}$ calculated between steps is equal to the step size divided by $q$, and can be visually compared with the periodic recurrence in linking number of discrete torque fluctuation states (Fig.~1b and Fig.~3b). We find that this model recapitulates the essential force- and ionic strength-dependent features of discontinuous states observed experimentally (Fig.~3), lending credence to the idea that torsionally buckled DNA has solenoidal superstructure in the force-extended region.

To investigate this further, a solenoid model predicts additional writhe not accounted for in a straight-rod model. This should have consequences for the torque in DNA prior to buckling. As seen in eqn.~\ref{eq:eqn1}, the torque, which is directly proportional to twist, is reduced by writhe and calculated using the effective torsional persistence length $C_{\rm eff}$. In particular, $C_{\rm eff}$ reflects the writhe ${\rm Wr}={\rm Lk}(1-C_{\rm eff}/C$) that develops as unbuckled DNA is overwound. The force dependence of this relationship is examined in Fig.~4: We compare the theory of torsional directed walks \cite{moroz} with the data of Mosconi {\it et al.} \cite{mosconi}. In their insightful theory, Moroz and Nelson derived the expression $C_{\rm eff}(f)\approx C(1+C/(4A\sqrt{\beta Af}))^{-1}$ by considering writhe caused by fluctuations about a straight-twisted rod \cite{moroz}. The straight-rod assumption may explain why, particularly at low forces, their expression overestimates the experimental values of $C_{\rm eff}$ (Fig.~4). This topic is highlighted by two alternative models recently proposed to explain the apparent discrepancy between the Moroz-Nelson theory and data \cite{nomidis,schurr}. Both models fit experimental data better than the Moroz-Nelson formula, but do so by including additional fitting parameters. Our model of solenoidal superstructure recasts the Moroz-Nelson result without additional parameters. Under the assumption that the total writhe includes linearly additive terms, we approximate a static component of writhe \cite{note5}, and find that a simple term, $-C(1-x_{\rm WLC}(f))$, augments the Moroz-Nelson theory and corrects the overestimate of $C_{\rm eff}$ remarkably well (Fig.~4). This suggests a modified formula,
\begin{equation}\label{eq:eqn7}
C_{\rm eff}(f)\approx(\frac{1}{C}+\frac{1}{4A\sqrt{\beta Af}})^{-1}-C(1-x_{\rm WLC}(f)),
\end{equation}
can be used as an improved estimate, which, when input to eqn.~\ref{eq:eqn1}, gives a reasonable approximation for the torque in overwound DNA \cite{SI,note4}.

The Moroz-Nelson theory is entirely self-consistent and the bare twist persistence length $C$ is correctly estimated using their procedure of fitting to $X$ vs ${\rm Lk}$ data \cite{moroz}. This works because the approximately quadratic change in extension used for fitting is caused by torque-coupled fluctuations. However, the discrepancy between the Moroz-Nelson formula for $C_{\rm eff}$ and data suggest there is an additional source of writhe that is decoupled from torque. We found that the missing writhe is consistent with a fluctuating solenoid -- rather than straight-rod -- shape (Fig.~4), and therefore recommend use of eqn.~\ref{eq:eqn7} as an improved approximation for $C_{\rm eff}$.

We presented experimental evidence that DNA supercoiling features unexpected discontinuities or steps occurring at highly regular intervals of extension or linking number, indicating sawtooth-like variations in torque associated with extending the plectoneme of buckled DNA. We found that both the observed steps and force-dependent torsional elasticity are quantitatively consistent with twisted and force-extended DNA storing writhe in the form of a solenoid structure. To show this, we compared our data to model predictions by developing a framework for computing the elastic response of DNA overwound at constant tension. To promote further measurements and theories relating shape to mechanics in DNA or other supercoiled biopolymers, we provide this as an open-source tool for general use \cite{note4}. Building from this template, our model might be refined and extended by exact calculations of fluctuating solenoid structure.


\begin{thebibliography}{10}
\bibitem{note1}
Data in Fig.~1b are data from Fig.~1a transformed by setting the average torque $\Gamma_b$ in the plectoneme equal to $2fq/(3\pi x_{\rm WLC})$ \cite{clauvelin}, with small changes in $\Gamma$ assumed linearly proportional to small changes in {\rm Lk}.
\bibitem{note4}
Our computational framework is further described and demonstrated in annotated Python code, which is appended here and will be available on GitHub.
\bibitem{note5}
The writhe in a static one-turn-per-wave solenoid is ${\rm Wr}={\rm Lk}(1-\sin{\gamma})$ \cite{plectenoid}, with solenoid opening angle $\gamma$. In our simple approximation, $\sin{\gamma}$ equals the wavelength divided by the contour length per wave, or fractional DNA extension, $x_{\rm WLC}$. This estimate avoids double counting of fluctuations already accounted for in the Moroz-Nelson theory \cite{note6}.
\bibitem{note2}
We posit $({\rm Lk}^\dagger-{\rm Lk}^*)\approx{\rm c_3}f/\Gamma_b(f,c)$, noting that it provides reasonable force dependence as a sublinear power-law, a logarithmic dependence on salt concentration, and works well over our specified parameter range \cite{SI,note4}. In our model, $({\rm Lk}^\dagger-{\rm Lk}^*)$ also determines the torque difference ($\Gamma^\dagger-\Gamma_b$) and is reflected in sigmoidal shape of the average extension vs excess linking number curve near buckling \cite{note3}.
\bibitem{note3}
In these probabilities, the argument of the exponential is $-\beta$ times the change in torsional free energy upon buckling, estimated by assuming a constant local torque drop, ($\Gamma^\dagger-\Gamma_b)=2\pi C_{\rm eff}/(\beta L_0)({\rm Lk}^\dagger-{\rm Lk}^*)$, for {\rm Lk} near ${\rm Lk}^\dagger$. This is consistent with the sigmoidal shape of the buckling transition in average elastic response curves, as well as experimental estimates of ($\Gamma^\dagger-\Gamma_b)=-(2\pi\beta)^{-1}\partial{\ln{\frac{P_u}{P_b}}}/\partial{{\rm Lk}}$ from two-state fluctuation data.
\end{thebibliography}
\end{document}